\documentclass[11pt,a4paper]{article}

\usepackage{epsfig}

\begin{document}

\title{\bf Continuum physics\\ with quenched overlap fermions}
\author{{\bf Stephan D\"urr}$^a$ and {\bf Christian Hoelbling}$^b$\\[2mm]
${}^a$Universit\"at Bern, ITP,
Sidlerstr.\,5, CH-3012 Bern, Switzerland\\
${}^b$Universit\"at\,Wuppertal,
Gaussstr.\,20, D-42119\,Wuppertal, Germany}
\date{}
\maketitle
\begin{abstract}
We calculate $m_{ud}\!=\!(m_u\!+\!m_d)/2$, $m_s$, $f_\pi$ and $f_K$ in the
quenched continuum limit with UV-filtered overlap fermions. We see rather small
scaling violations on lattices as coarse as $a^{-1}\!\simeq\!1\,\mathrm{GeV}$
and conjecture that similar advantages would be manifest in unquenched studies.
\end{abstract}

\newcommand{\pad}{\partial}
\newcommand{\ovr}{\over}
\newcommand{\til}{\tilde}
\newcommand{\pri}{^\prime}
\renewcommand{\dag}{^\dagger}
\newcommand{\<}{\langle}
\renewcommand{\>}{\rangle}
\newcommand{\nab}{\nabla}
\newcommand{\gaf}{{\gamma_5}}

\newcommand{\al}{\alpha}
\newcommand{\be}{\beta}
\newcommand{\ga}{\gamma}
\newcommand{\de}{\delta}
\newcommand{\ep}{\epsilon}
\newcommand{\ve}{\varepsilon}
\newcommand{\ze}{\zeta}
\newcommand{\et}{\eta}
\renewcommand{\th}{\theta}
\newcommand{\vt}{\vartheta}
\newcommand{\io}{\iota}
\newcommand{\ka}{\kappa}
\newcommand{\la}{\lambda}
\newcommand{\rh}{\rho}
\newcommand{\vr}{\varrho}
\newcommand{\si}{\sigma}
\newcommand{\ta}{\tau}
\newcommand{\ph}{\phi}
\newcommand{\vp}{\varphi}
\newcommand{\ch}{\chi}
\newcommand{\ps}{\psi}
\newcommand{\om}{\omega}

\newcommand{\bdm}{\begin{displaymath}}
\newcommand{\edm}{\end{displaymath}}
\newcommand{\bea}{\begin{eqnarray}}
\newcommand{\eea}{\end{eqnarray}}
\newcommand{\beq}{\begin{equation}}
\newcommand{\eeq}{\end{equation}}

\newcommand{\mr}{\mathrm}
\newcommand{\mb}{\mathbf}
\newcommand{\Nf}{{N_{\!f}}}
\newcommand{\Nc}{{N_{\!c}}}
\newcommand{\ri}{\mr{i}}
\newcommand{\MeV}{\,\mr{MeV}}
\newcommand{\GeV}{\,\mr{GeV}}
\newcommand{\fm}{\,\mr{fm}}


\section{Introduction}

Overlap fermions \cite{Neuberger:1997fp} satisfy the Ginsparg-Wilson relation
\cite{Ginsparg:1981bj} 
\beq
\gaf D+D\hat\gaf=0\;,\qquad\hat\gaf=\gaf(1-{1\over\rho}D)
\label{GW}
\eeq
which implies exact chiral symmetry at finite lattice spacing
\cite{Hasenfratz:1998ri,Luscher:1998pq}.
While theoretically clean, calculations with overlap fermions are considered
a computational challenge.
In a recent investigation \cite{Durr:2005an} it has been conjectured that a
UV-filtered (smeared) Wilson kernel operator, as previously suggested in
\cite{Bietenholz:2000iy,Kovacs:2001bx,Kovacs:2002nz,DeGrand:2002va}, might
substantially reduce the computational cost associated with obtaining
continuum physics \cite{Dong:2000mr}.
In \cite{Durr:2005an} the focus was on technical aspects, but it is clear
that the point relevant in practical applications is whether UV-filtered
(``thick link'') overlap fermions would extend
the scaling region to significantly coarser lattices, even if one refrains
from ($\be$-specific) tuning and sets $\rh\!=\!1$.
The present note addresses this question by studying the continuum limit of
the pseudoscalar masses and decay constants with quenched UV-filtered overlap
quarks.
No systematic comparison to plain (``thin link'') overlap fermions is made,
because we could not obtain reasonable signals for plain overlap
fermions with $\rh\!=\!1$ on our coarser lattices.
Given the exploratory nature of the present investigation, we restrict
ourselves to pseudoscalar meson correlators in the $p$-regime of Chiral
Perturbation Theory (XPT) \cite{Gasser:1983yg}.
Our main results are the values of the quenched strange quark mass and $K$
decay constant in the continuum
\beq
m_s(\overline{\mr{MS}},2\GeV)=119(10)(7)\MeV
\;,\qquad
f_K=170(10)(2)\MeV
\label{results_one}
\eeq
as well as the quark mass and decay constant ratios
\beq
{m_s\over m_{ud}}=23.3(7.1)(4.5)
\;,\qquad
{f_K/f_\pi}=1.17(4)(2)
\label{results_two}
\eeq
where $m_{ud}\!=\!(m_u\!+\!m_d)/2$ and no numerical input from XPT has been
used.
Throughout this note the first error is statistical and the second systematic,
but the latter does \emph{not} include any estimate of the quenching effect.


\section{Technical setup}

We use the Wilson gauge action.
The massless overlap operator is \cite{Neuberger:1997fp}
\beq
D_\mr{ov}=\rho\Big[
1+(D_\mr{W}-\rho)\big((D\dag_\mr{W}-\rho)(D_\mr{W}-\rho)\big)^{-1/2}
\Big]
\label{def_Dzero}
\eeq
with $D_\mr{W}$ the massless Wilson operator.
The UV-filtered overlap is construc\-ted by evaluating the Wilson operator on
APE \cite{Albanese:ds} or HYP \cite{Hasenfratz:2001hp,Hasenfratz:2001tw}
smeared gauge configurations, resulting in an $O(a^2)$ redefinition of the
fermion action \cite{Durr:2004as,Durr:2005an}.
We use 1 or 3 iterations with smearing parameters $\alpha_\mr{APE}\!=\!0.5$ or
$\alpha_\mr{HYP}\!=\!(0.75,0.6,0.3)$ and the shift parameter $\rho\!=\!1$ is
kept fixed.
Based on (\ref{def_Dzero}) the massive operator is defined through
\beq
D_{\mr{ov},m}=\Big(1-{am\ovr2\rh}\Big)D_\mr{ov}+m
\;.
\label{def_Dmass}
\eeq

We set the lattice spacing with the Sommer parameter \cite{Sommer:1993ce},
i.e.\ we give all intermediate results in appropriate powers of $r_0$ to
facilitate comparison with other quenched studies.
Only the final result will be converted into physical units assuming a
standard value for $r_0$.
We have data at 4 different lattice spacings in matched boxes of physical
volume $L^3\!\times\!2L$ with $L\!\simeq\!3r_0$.
The couplings were chosen with the interpolation formula given in
\cite{Guagnelli:1998ud}.
The details of the simulation are summarized in Tab.\,\ref{tab_param}.

\begin{table}[!b]
{}\hfill
\begin{tabular}{c|cccc}
$N_L^3\!\times\!N_T^{}$&
$8^3\!\times\!16$&$10^3\!\times\!20$&$12^3\!\times\!24$&$16^3\!\times\!32$\\
\hline
$\beta$& 5.66& 5.76& 5.84& 6.0 \\
$a$[fm]&0.188&0.149&0.125&0.093\\
\#conf.&  30 &  30 &  30 &  30
\end{tabular}
\hfill{}
\caption{Simulation parameters and statistics. The box volume in physical units
$L^3\!\times\!T\!=\!(1.5\fm)^3\!\times\!3.0\fm$ (based on $r_0\!=\!0.5\fm$) is
kept fixed.}
\label{tab_param}
\end{table}

For each coupling and filtering level we use 4 bare quark masses.
Ideally one would choose them such as to always obtain the same 4 renormalized
quark masses in $r_0$ units (or the same 4 pseudoscalar masses), but for this
one would have to know the renormalization factor $Z_m\!=\!Z_S^{-1}$
beforehand.
Our values as summarized in Tab.\,\ref{tab_mq} are not bad; our renormalized
masses are roughly in the region ${1\ovr3}m_s^\mr{phys}\!\ldots\!m_s^\mr{phys}$.

\begin{table}[!t]
{}\hfill
\begin{tabular}{c|llll}
&$\;\;\; 8^3\times16$&$\;\;\;10^3\times20$
&$\;\;\;12^3\times24$&$\;\;\;16^3\times32$\\
\hline
none   & 0.16 ... 0.4 & 0.16 ... 0.4 & 0.16 ... 0.4   & 0.08 ... 0.2  \\
1\,APE & 0.08 ... 0.2 & 0.08 ... 0.2 & 0.04 ... 0.1   & 0.04 ... 0.1  \\
3\,APE & 0.04 ... 0.1 & 0.04 ... 0.1 & 0.03 ... 0.075 & 0.03 ... 0.075\\
1\,HYP & 0.08 ... 0.2 & 0.08 ... 0.2 & 0.04 ... 0.1   & 0.04 ... 0.1  \\
3\,HYP & 0.04 ... 0.1 & 0.04 ... 0.1 & 0.03 ... 0.075 & 0.03 ... 0.075
\end{tabular}
\hfill{}
\caption{The 4 regularly spaced bare quark masses per coupling and filtering.}
\label{tab_mq}
\end{table}

In the course of this calculation we will need both $Z_S\!=\!Z_P$ (to determine
the renormalized quark masses) and $Z_V\!=\!Z_A$ (for the decay constants),
where the alleged identity is specific for the massless overlap operator.

\begin{figure}[!b]
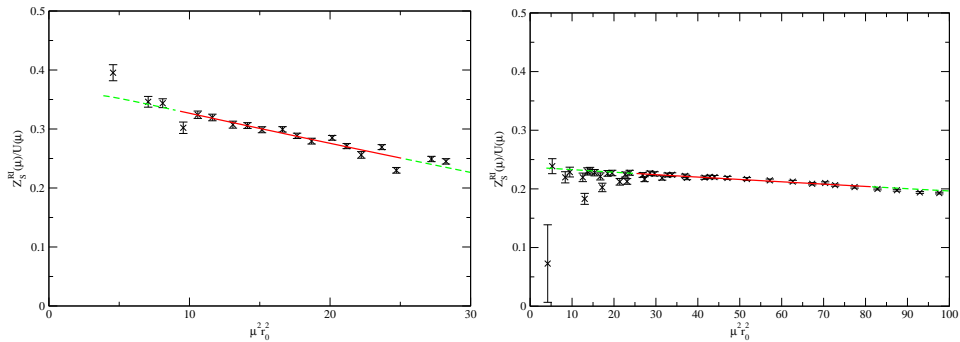

\epsfig{file=zs_hyp1_08.eps,width=6.25cm}
\epsfig{file=zs_hyp1_16.eps,width=6.25cm}
\vspace*{-6mm}
\caption{$Z_S^\mr{RI-MOM}(\mu)/U(\mu)$ on our coarsest and finest
($\be\!=\!5.66,6.0$) lattice using the 1\,HYP overlap operator with
$\rh\!=\!1$. The solid line indicates a linear fit with range
$3\!\leq\!\mu r_0\!\leq\!5$ and $5\!\leq\!\mu r_0\!\leq\!9$, respectively.}
\label{fig_zs}
\end{figure}

To compute the scalar and pseudoscalar renormalization constant we follow the
non-perturbative RI-MOM procedure as defined in \cite{Martinelli:1994ty} and
first applied to overlap fermions in \cite{Giusti:2001yw,Giusti:2001pk}.
We compute both $Z_S(\mu)$ and $Z_P(\mu)$; in the latter case a $1/m$ term is
used to extrapolate the result to the chiral limit.
It turns out that the values obtained are in good agreement even on our
coarsest lattice.
We fit the scalar renormalization constant to the form
\beq
Z_S^\mr{RI-MOM}(\mu)=U(\mu)Z_S^{\mr{RGI}}+\mr{const}\,(a\mu)^2
\label{eq_zsfit}
\eeq
where $U(\mu)$ is the 4-loop running in the RI-MOM scheme as given in
\cite{Chetyrkin:1999pq} and the second term is introduced to account for
discretization effects.
In other words the scalar renormalization constant after dividing out the
4-loop perturbative running should be flat, up to discretization effects,
and Fig.\,\ref{fig_zs} shows that the latter are indeed non-negligible.
The wiggles in the data signal rotational symmetry breaking on the lattice.
We checked that within errors the slope disappears in proportion to
$(a/r_0)^2$.
The phenomenological analysis below is based on $Z_S^\mr{RI-MOM}(2\GeV)$,
where $\mu\!=\!2\GeV$ is realized through $\mu r_0\!=\!5.06773$.
The systematic error is estimated by varying the fit range, by
including additional $1/p^2$ terms into the fit, and by comparing to the
$Z_P^\mr{RI-MOM}(2\GeV)$ data.
A summary of our results, after conversion to $(\overline{\mr{MS}},2\GeV)$
conventions, is presented in Tab.\,\ref{tab_zs}.
Choosing a fixed $\rh\!>\!1$ would delay the breakdown of the unfiltered
version -- see \cite{Durr:2005an} for details.
Note that the $Z_S$ factors of all UV-filtered operators are much closer to
$1$, even when compared to the unfiltered overlap operator with tuned $\rh$
\cite{Giusti:2001yw,Hernandez:2001yn,Wennekers:2005wa}.
This suggests that one should be able to compute renormalization constants
perturbatively, as was done in \cite{DeGrand:2002va} in a slightly different
setup.

\begin{table}[!t]
{}\hfill
\begin{tabular}{c|cccc}
 & $\be=5.66$ & $\be=5.76$ & $\be=5.84$ & $\be=6.0$ \\
\hline
none   & ill-def.    & ill-def.   & 6.14(15)(38) & 2.54(6)(12) \\
1\,APE & 3.05(7)(34) & 1.79(4)(6) & 1.60(3)(8)   & 1.26(3)(8) \\
3\,APE & 2.05(7)(43) & 1.35(4)(8) & 1.25(2)(8)   & 1.04(2)(7) \\
1\,HYP & 1.71(4)(24) & 1.24(3)(6) & 1.23(2)(6)   & 1.03(2)(7) \\
3\,HYP & 1.57(6)(25) & 1.16(3)(5) & 1.10(2)(7)   & 0.98(2)(6)
\end{tabular}
\hfill{}
\caption{Renormalization constant $Z_S^{\overline{\mr{MS}}}(2\GeV)$ for Wilson
glue and $\rh\!=\!1$. The main effect of filtering is to bring it much closer
to its tree-level value~1.}
\label{tab_zs}
\end{table} 


The second ingredient is the axial-vector renormalization constant $Z_A$.
Here we use the values given in \cite{Durr:2005an} (coming from a PCAC
renormalization condition), complemented by a $\be\!=\!5.76$ column included
in \cite{DuHoWe_proceedings}.
It turns out that the values in this column are in fair agreement
with the prediction by the Pad\'e curve given in \cite{Durr:2005an},
which is another indication that for the filtered overlap operator lattice
perturbation theory might work rather well.


\section{Physical results}

To extract meson masses and decay constants we compute the correlators
\beq
C_{\Gamma_1,\Gamma_2}(t)=\sum_{\mb{x}}
\<\bar\psi_1(0)\Gamma_1\hat\psi_2(0)
\bar\psi_2(\mb{x},t)\Gamma_2\hat\psi_1(\mb{x},t)\>
\label{eq_corr}
\eeq
where
\beq
\hat{\psi}={1+\gaf\hat{\gaf}\over 2}\psi
\eeq
denotes the ``chirally rotated'' quark field \cite{Luscher:1998pq}.
Specifically, we consider
\beq
\<SS\>(t)=C_{1,1}(t)
\;,\quad
\<PP\>(t)=C_{\gaf,\gaf}(t)
\;,\quad
\<A_0A_0\>(t)=C_{\gaf\gamma_0,\gaf\gamma_0}(t)
\eeq
and we extract the mass and decay constant of the pseudoscalar meson from
fits to the $\<A_0A_0\>$ and $\<PP\>\!-\!\<SS\>$ \cite{Blum:2000kn} channels.
We do not consider $\<PP\>$, since it is known to be contaminated
by zero mode contributions \cite{Giusti:2001pk,Gattringer:2003qx}.

Generically, we see plateaus in the effective mass from about $N_T/4$ on, with
the exception of the unfiltered operator, where the plateau sets in later and
is less pronounced.
Our central values stem from a fit to the $\<A_0A_0\>$ correlator in the range
$5 N_T/16\!\le\!t\!\le\!11N_T/16$.
The theoretical error is dominated by the comparison to $\<PP\>\!-\!\<SS\>$.
Including only variations of the fit range (and below the error on $Z_S$),
it would be much smaller.

\begin{figure}[!t]
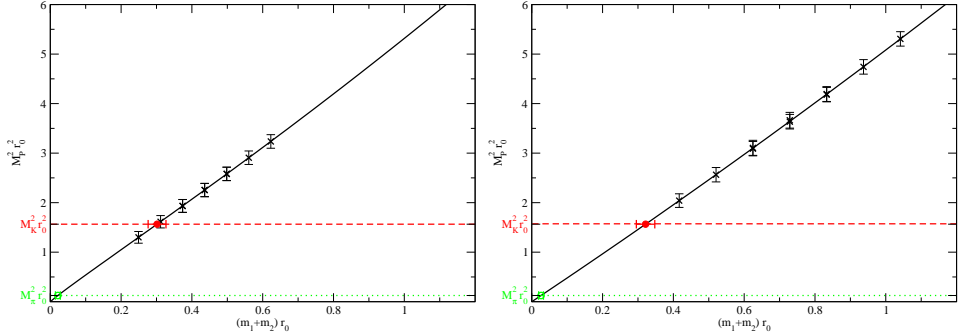

\epsfig{file=mpi_hyp1_08.eps,width=6.25cm} 
\epsfig{file=mpi_hyp1_16.eps,width=6.25cm} 
\vspace*{-6mm}
\caption{$M_Pr_0$ versus the bare quark mass $(m_1\!+\!m_2)r_0$ on our coarsest
and finest ($\be\!=\!5.66,6.0)$ lattice for the 1\,HYP operator. The masses
come from a fit to the $\<A_0A_0\>$ correlator in the range
$5N_T/16\!\le\!t\!\le\!11N_T/16$. The solid curve represents a fit to the
functional form (\ref{eq_chilogfit}) and the horizontal lines indicate the
physical $M_\pi r_0$ and $M_K r_0$ values. The latter are used to read off
the fitted $2m_{ud}r_0$ and $(m_s\!+\!m_{ud})r_0$, respectively.}
\label{fig_chilogfit}
\end{figure}

In Fig.\,\ref{fig_chilogfit} the pseudoscalar meson mass squared is plotted
versus the bare quark mass.
Using regularly spaced quark masses, each $m_1\!+\!m_2$ combination is realized
in several ways and the pertinent $M_P$ are in excellent agreement.
In other words, isospin breaking effects are completely negligible.
Therefore we fit the quark mass dependence of the pseudoscalar meson mass
with the resummed quenched XPT expression \cite{Sharpe:1992ft}
\beq
M_P^2=A\,(m_1\!+\!m_2)^{1/(1+\de)}+B\,(m_1\!+\!m_2)^2
\label{eq_chilogfit}
\eeq
which is strictly true only for degenerate masses.
Our $\de$ fluctuates wildly and is thus trivially consistent with 0.2
\cite{Sharpe:1992ft}.

The next step requires some experimental input.
Since we do not see any isospin breaking effects and electromagnetic
corrections are of the order of 1\%, while our statistical errors turn out to
be roughly 10\%, we decided to use the charged meson masses as input.
Sticking to the identification $r_0\!=\!0.5\fm$ we use
$M_Kr_0\!=\!493.7\MeV\!\cdot0.5\fm\!=\!1.251$ and
$M_\pi r_0\!=\!139.6\MeV\!\cdot0.5\fm\!=\!0.3537$ as input.
This gives the bare $(m_1\!+\!m_2)r_0$ values indicated by a horizontal error
bar in Fig.\,\ref{fig_chilogfit}.
Multiplying the bare masses with the $Z_m\!=\!Z_S^{-1}$ obtained earlier,
we extract the renormalized quark masses.
Our results for $m_s\!+\!m_{ud}$ and $2m_{ud}$ are reported in
Tabs.\,\ref{tab_ms} and \ref{tab_mud}, respectively.

\begin{table}
{}\hfill
\begin{tabular}{c|cccc}
 & $\be=5.66$ & $\be=5.76$ & $\be=5.84$ & $\be=6.0$ \\
\hline
none   & ill-def.      & ill-def.      & 0.219(34)(181)& 0.378(36)(28)\\
1\,APE & 0.244(34)(45) & 0.366(40)(50) & 0.260(33)(19) & 0.329(27)(31)\\
3\,APE & 0.260(21)(47) & 0.311(32)(43) & 0.223(32)(20) & 0.312(26)(26)\\
1\,HYP & 0.301(25)(39) & 0.336(35)(67) & 0.240(28)(23) & 0.321(26)(28)\\
3\,HYP & 0.281(20)(40) & 0.321(31)(24) & 0.247(31)(21) & 0.321(26)(22)
\end{tabular}
\hfill{}
\caption{$(m_s\!+\!m_{ud})r_0$ in $(\overline{\mr{MS}},2\GeV)$ conventions. For
the $\rh\!=\!1$ thin link action at $\be\!=\!5.66,5.76$ no $Z_m\!=\!Z_S^{-1}$
is available (cf.\ Tab.\,3).}
\label{tab_ms}
\vspace*{4mm}
{}\hfill
\begin{tabular}{c|cccr}
 & $\be=5.66$ & $\be=5.76$ & $\be=5.84$ & $\be=6.0$ \\
\hline
none   & ill-def.      & ill-def.      & 0.006(05)(08) & 0.027(09)(03)\\
1\,APE & 0.006(07)(09) & 0.017(12)(13) & 0.007(08)(04) & 0.027(07)(04)\\
3\,APE & 0.012(05)(17) & 0.014(10)(18) & 0.007(07)(01) & 0.023(06)(03)\\
1\,HYP & 0.021(07)(05) & 0.021(08)(09) & 0.011(07)(02) & 0.026(07)(04)\\
3\,HYP & 0.014(08)(05) & 0.028(12)(09) & 0.009(06)(05) & 0.022(05)(03)
\end{tabular}
\hfill{}
\caption{$2m_{ud}r_0$ in $(\overline{\mr{MS}},2\GeV)$ conventions. For
the $\rh\!=\!1$ thin link action at $\be\!=\!5.66,5.76$ no $Z_m\!=\!Z_S^{-1}$
is available (cf.\ Tab.\,3).}
\label{tab_mud}
\end{table}
\begin{figure}
\epsfig{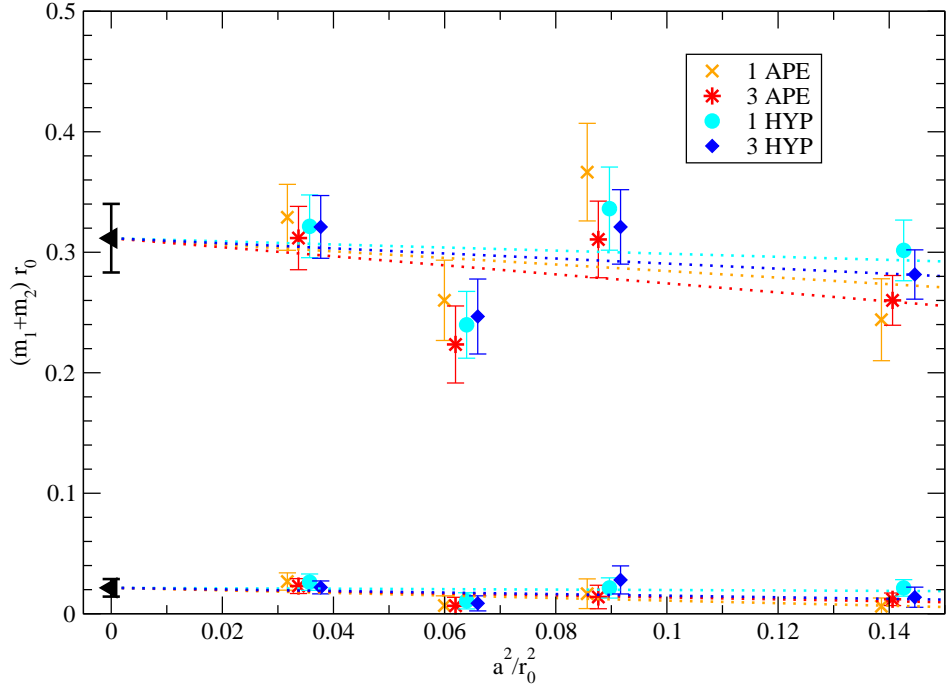} 
\vspace*{-4mm}
\caption{Renormalized $(\overline{\mr{MS}},2\GeV)$ masses $(m_s\!+\!m_{ud})r_0$
and $2m_{ud}r_0$ versus $(a/r_0)^2$. Errors are statistical, the 4 filterings
are correlated (same configs.).}
\label{fig_ctfit}
\end{figure} 

To finally push to the continuum we perform a combined fit (including all
couplings and smearing levels) with a common continuum value and individual
$O(a^2)$ terms.
We do this for the quantities $(m_s\!+\!m_{ud})r_0$, $2m_{ud}r_0$
(as shown in Fig.\,\ref{fig_ctfit}) as well as for $m_sr_0$ and $m_s/m_{ud}$.
Our final result is
\beq
m_s(\overline{\mr{MS}},2\GeV)r_0=0.301(25)(7)
\eeq
in the continuum [$\ch^2\!/\mr{d.o.f.}\!=\!2.1$] which, upon using
$r_0\!=\!0.5\fm$, leads to the result given in (\ref{results_one}).
For the ratio $m_s/m_{ud}$ we find the value quoted in (\ref{results_two})
[$\ch^2\!/\mr{d.o.f.}\!=\!0.32$].
In either case the generic comment on statistical and systematic errors, as
stated above, applies.
Evidently, any result involving $m_{ud}$ benefits from our choice to use
(\ref{eq_chilogfit}) at finite lattice spacing, since it restricts the curve
in Fig.\,\ref{fig_chilogfit} to go through zero.
The absence of curvature in our data is the reason why this ratio is in
rather good agreement with the XPT result $m_s/m_{ud}\!=\!24.4(1.5)$
\cite{Gasser:1982ap}, in spite of the limitations mentioned and in spite
of the latter value referring to a different theory (full QCD).

\begin{figure}[!t]
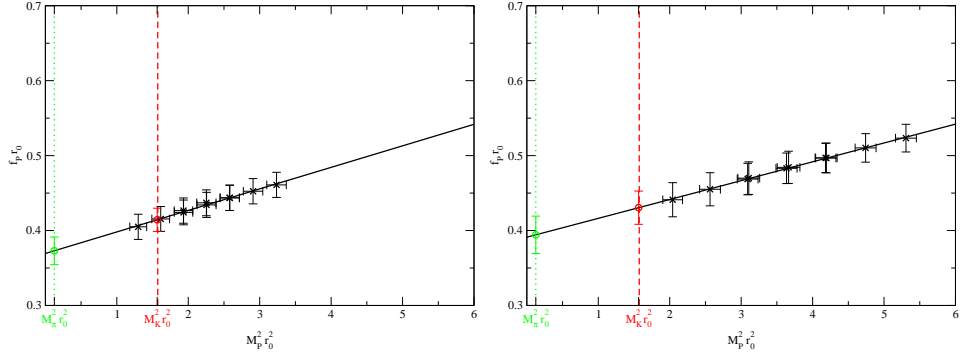

\epsfig{file=fpi_hyp1_08.eps,width=6.25cm} 
\epsfig{file=fpi_hyp1_16.eps,width=6.25cm} 
\vspace*{-6mm}
\caption{Pseudoscalar decay constant $f_Pr_0$ versus $(M_Pr_0)^2$ on our
coarsest and finest $(\be=5.66,6.0)$ lattice for the 1\,HYP overlap operator.
$M_P$ and $f_P$ are obtained from a fit to the $\<A_0A_0\>$ correlator in
the region $T=\{5 N_T/16,\ldots,11 N_T/16\}$. The solid curve is a linear
fit, the dotted vertical lines indicate the physical $M_\pi r_0$ and $M_K r_0$,
respectively.}
\label{fig_fpfit}
\end{figure} 

The pseudoscalar decay constant $f_P$ is extracted from the $\<A_0A_0\>$
correlator.
Generically, we see plateaus for the effective decay constant from about
$N_T/4$ on, again with the exception of the unfiltered operator where they
are less pronounced.
We use the $Z_A$ values discussed earlier.
Our central values stem from the interval $5 N_T/16\!\le\!t\!\le\!11N_T/16$
and the error estimate comes from comparing to the $\<PP\>\!-\!\<SS\>$ channel,
from varying the fit range and from the error on $Z_A$.

\begin{table}
\begin{tabular}{c|cccc}
 & $\be=5.66$ & $\be=5.76$ & $\be=5.84$ & $\be=6.0$ \\
\hline
none   & ill-def.      & ill-def.      & 0.575(39)(163) & 0.422(19)(10) \\
1\,APE & 0.412(23)(69) & 0.391(21)(10) & 0.463(31)(35)  & 0.430(22)(26) \\
3\,APE & 0.426(18)(54) & 0.398(20)(06) & 0.449(33)(33)  & 0.433(22)(31) \\
1\,HYP & 0.414(15)(40) & 0.398(21)(03) & 0.446(32)(22)  & 0.430(22)(27) \\
3\,HYP & 0.400(15)(49) & 0.378(16)(08) & 0.419(28)(26)  & 0.430(22)(35)
\end{tabular}
\caption{$f_K r_0$ on all lattices used in the continuum extrapolation and
the unfiltered. With the $\rh\!=\!1$ thin link action at $\be\!=\!5.66,5.76$
there is no signal.}
\label{tab_fk}
\vspace*{4mm}
\begin{tabular}{c|cccc}
 & $\be=5.66$ & $\be=5.76$ & $\be=5.84$ & $\be=6.0$ \\
\hline
none   & ill-def.       & ill-def.      & 0.541(48)(211) & 0.387(23)(19) \\
1\,APE & 0.367(37)(161) & 0.356(24)(22) & 0.438(45)(81)  & 0.393(25)(36) \\
3\,APE & 0.398(36)(133) & 0.358(24)(28) & 0.415(47)(66)  & 0.397(26)(42) \\
1\,HYP & 0.373(18)(69)  & 0.365(24)(14) & 0.412(42)(33)  & 0.394(25)(36) \\
3\,HYP & 0.386(26)(118) & 0.334(18)(07) & 0.386(39)(55)  & 0.396(25)(45) 
\end{tabular} 
\caption{$f_\pi r_0$ on all lattices used in the continuum extrapolation and
the unfiltered. With the $\rh\!=\!1$ thin link action at $\be\!=\!5.66,5.76$
there is no signal.}
\label{tab_fpi}
\end{table}
\begin{figure}
\epsfig{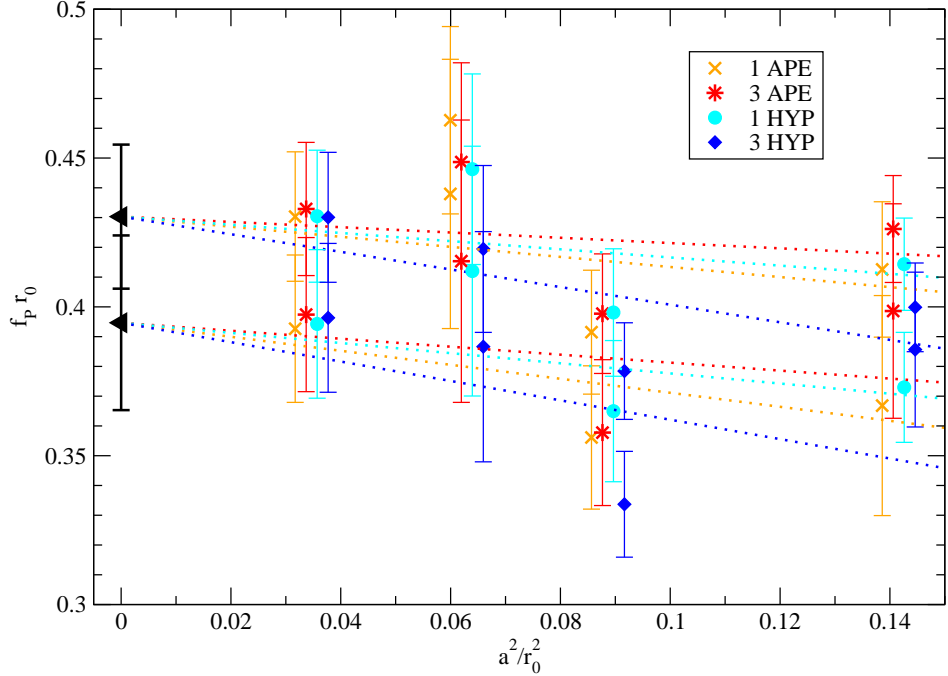} 
\vspace*{-4mm}
\caption{Pseudoscalar decay constants $f_\pi r_0$ and $f_K r_0$ versus
$(a/r_0)^2$. Errors are statistical, the 4 filterings are correlated (same
configs.).}
\label{fig_ctffit}
\end{figure} 

In Fig.\,\ref{fig_fpfit} $f_P$ is plotted versus the meson mass squared.
We see a strictly linear behavior at all couplings, with the lattice spacing
having only mild effects on intercept and slope.
Since (quenched) XPT predicts a linear dependence on $M_P^2$ at lowest order
and (quenched) chiral logs enter at 1-loop level only, we decided to stick to
the leading order and do a linear fit.
The intercept with $(M_Kr_0)^2\!=\!1.251^2$ and $(M_\pi r_0)^2\!=\!0.3537^2$
defines the $f_Kr_0$ and $f_\pi r_0$ at finite lattice spacing reported in
Tabs.\,\ref{tab_fk} and \ref{tab_fpi}, respectively.

To extrapolate to the continuum, the same recipe is used as before.
A combined fit to all filtering levels with a common continuum limit and
individual $O(a^2)$ terms is shown in Fig.\,\ref{fig_ctffit} for $f_K$ and
$f_\pi$.
We obtain
\beq
f_K r_0=0.430(24)(5)
\eeq
in the continuum [$\ch^2\!/\mr{d.o.f.}\!=\!0.96$] which, via $r_0\!=\!0.5\fm$,
yields (\ref{results_one}).
Applying the same strategy to $f_K/f_\pi$ we find the value quoted
in (\ref{results_two}) [$\ch^2\!/\mr{d.o.f.}\!=\!0.52$].
Given the large errors it is not so surprising that both $f_K$ and
$f_K/f_\pi$ are in reasonable agreement with the experimental
$f^\mr{exp}_K=159.8(1.5)\MeV$ and $f_K^\mr{exp}/f_\pi^\mr{exp}=1.223(12)$,
respectively \cite{Hagiwara:2002fs}.
Especially in the latter case, higher precision would likely reveal a genuine
quenching error \cite{Aoki:2002fd}.


\section{Conclusion}

We have presented an exploratory scaling study with UV-filtered overlap
quarks, concentrating on the quenched continuum limit of the light quark masses
and the $\pi$ and $K$ decay constants.
The focus is not so much on the numerical values obtained; the point is that
a nice scaling behavior has been observed down to lattices as
coarse as $a^{-1}\!\simeq\!1\GeV$.
As anticipated in \cite{Durr:2005an} the details of the filtering recipe play
a minor role and therefore we conjecture that with analytic gauge link
smearing techniques \cite{Morningstar:2003gk} these properties will carry over
to dynamical overlap calculations
(cf.\ \cite{DeGrand:2004nq,DeGrand:2005vb,kalman}).

For the time being let us mention that the simulations were done on a
few of todays standard PCs (3\,GHz Pentium\,4), using an equivalent
of $\sim\!120$ days on one such PC for the 1\,APE operator (including all
couplings) and of $\sim\!60$ days for any of the other filtered operators.
With a complete scaling study based on overlap valence quarks requiring less
than 1 PC year one is tempted to ask whether for some phenomenological
questions the overlap might actually represent the cheapest approach
to obtain a continuum result.

\bigskip

\noindent
{\bf Acknowledgments}: This work is based on an earlier collaboration with Urs
Wenger that lead to \cite{Durr:2005an}. S.D.\ was supported by the Swiss NSF.


\end{document}